\documentclass[aps,twocolumn,prl,showpacs]{revtex4}
\usepackage{amssymb}
\usepackage{epsfig}
\usepackage{graphicx}
\usepackage{amssymb}
\usepackage{wasysym}

\begin{document}

\title{Autocatalytic chemical smoke rings}

\author{Michael C. Rogers and Stephen W.
Morris}
\affiliation{Department of Physics, University of
Toronto, 60 St. George St., Toronto, Ontario, Canada M5S 1A7}

\date{\today}

\begin{abstract}

Buoyant plumes, evolving free of boundary constraints, may develop 
well-defined mushroom shaped heads. In normal plumes, overturning 
flow in the head entrains less buoyant fluid from the surroundings as the head rises, 
robbing the plume of its driving force.  We consider here a new type
of plume in which the source of buoyancy is an autocatalytic chemical 
reaction. The reaction occurs at a sharp front which separates reactants from less 
dense products.  In this type of plume, entrainment assists the reaction, producing new 
buoyancy which fuels an accelerating plume head. When the head has grown to a 
critical size, it detaches from the upwelling conduit, forming an accelerating, buoyant 
vortex ring.  This  vortex is analogous to a  rising smoke ring. 
 A second-generation head then develops at the point of detachment.
Multiple generations of chemical vortex rings can detach from a single 
triggering event.

\end{abstract}


\maketitle

Hydrodynamic flows driven by chemical reactions have been studied since the
discovery of fire, the oldest technology of humankind~\cite{combust}.  
Exothermic combustion reactions produce buoyant flames in the form of rising,
reacting plumes and thermals which are often highly turbulent~\cite{fires}.  In this Letter we examine
a much simpler, yet still very rich, new system in which relatively gentle 
autocatalytic reaction fronts give rise to laminar plumes and thermals in the form of free vortex rings. 
Unlike conventional non-reacting plumes and thermals arising from localized
heat sources, these flows have an internal source of buoyancy which is enhanced by subsequent  entrainment, leading to flow acceleration similar to combustion.  They provide a new and especially elegant model system in which to study instabilities in reacting flows.

Many autocatalytic chemical reactions exhibit sharp fronts.  The well-studied reaction we employed was the iodate oxidation of arsenous acid~\cite{hanna}. The short range diffusion of the autocatalyst, here the iodide ion, limits 
the reaction to a very thin front separating reacted and unreacted solution. 
The front region of active reaction is typically only a few tens of microns thick, 
and is by far the smallest lengthscale in the system.
Under gravity, hydrodynamic convection can be driven by buoyancy due to the density jump 
across the front. Density change is created both by thermal expansion due to the 
slight exothermicity of the reaction, and by the partial molal density decrease of 
the product solution~\cite{pojman}. Ascending fronts are hydrodynamically unstable because they put less dense products below denser unreacted solution~\cite{hanna,pojman,vasquez_01,maser,wilder,huang,carey,bockmann,dewit, edwards, leconte}.

In our plume experiments, we increase the viscosity 
of the water solution by adding glycerol. An ascending front is allowed to escape from
a capillary tube into a much larger volume, so that boundaries no longer
constrain the shape of the front.  In the larger volume, a rising plume with
a well-defined head is formed. During the early stages of the evolution of the
head, its growth is due to both a volume flux
of upwelling product solution and by the entrainment of reactant solution.
Later, the plume head detaches and becomes a rising vortex ring, analogous to a
chemically reacting smoke ring.  The upwelling conduit that is left behind
forms a new head which also eventually detaches, leading to generations of
accelerating vortex rings.  This surprising behaviour is in contrast to conventional buoyant vortex rings,
which do not accelerate.  It may have application as a laboratory analog of plumes in internally heated fluids, such as the Earth's mantle~\cite{mantle, griffiths_01}.

%
%
The propagation of iodate-arsenous acid (IAA)  reaction fronts has enjoyed
considerable theoretical
and experimental attention. It has been well-studied in capillary
tubes~\cite{pojman,vasquez_01,maser}, in thin 
slots~\cite{wilder,huang,carey,bockmann,dewit},
and in the presence of a superposed
flow~\cite{edwards, leconte}.  In sufficiently narrow vertically oriented capillary
tubes, convection is suppressed and the sharp reaction front is observed to
propagate at constant speed in either direction.  If the tube radius $a$ is
increased beyond a critical size, the front deforms and upward propagation
is accompanied by convection~\cite{vasquez_02}.  Within a simple thin front
approximation~\cite{edwards}, the relative importance of buoyancy is
described by a dimensionless quantity
\begin{equation}
S=\frac{\delta g a^3}{\nu D_c},
\end{equation}
where $g$ is the acceleration due to gravity, $\nu$  is the kinematic
viscosity, $D_c$ is the diffusion constant of the autocatalyst, and
$\delta=(\rho_u/\rho_r)-1$  is the dimensionless density jump
between the reacted ($\rho_r$) and unreacted ($\rho_u$) solutions.
In practice, for the IAA reaction, $\delta \sim 10^{-4}$.  The critical tube
radius for convection corresponds to a critical value $S_c \sim 90$.  At
slightly larger values $S > S_c$, a transition to an axisymmetric mode of
convection can be identified~\cite{pojman}. In the plume experiments 
described below, a very large tube was used, corresponding to $S \approx 10^{8}$.

In all previous studies of the IAA reaction~\cite{hanna, pojman, vasquez_01, maser, wilder, huang, carey, bockmann, dewit, edwards, leconte}, convection effects were severely
constrained by the viscous interaction with nearby solid boundaries.  We set out to find the reaction-driven flow phenomenology for an {\it unbounded} solution in which the reaction is initiated at a point, as nearly as possible.  Two experimental difficulties immediately arise. For a water solution, the IAA reaction produces fronts that are
sufficiently buoyant that the resulting plumes ascend quickly to the upper 
boundaries of any tank of managable size.  These plumes also had Reynolds numbers, $Re \sim 7$ 
and therefore had a rather complex internal
flow structure.  We therefore explored the reaction in a water-glycerol solution in which the increased viscosity acts to limit the
Reynolds number to $Re \sim 0.1$,  and to allow for an effectively longer evolution time within the experimental tank. These experiments reveal plumes with well-defined heads which eventually pinch-off to form essentially free vortex rings.

The second difficulty concerns initiating the front at a localized point.  When the front is released from a very narrow capillary tube with diameter $0.9$  mm into a much larger volume, we found that the front did not propagate.  This ``front death" phenomenon~\cite{front_death} is apparently due to the requirement of a minimum critical concentration of autocatalyst near the initiation region, which therefore cannot be too small.  To our knowledge, this effect has not been observed in the IAA reaction before.  This constraint is easily overcome by using a large enough capillary tube  to launch the reaction into the unbounded region. A diameter of $ 2.7$ mm proved to be sufficient.

%
%
Reactant solutions were prepared using
reagent grade chemicals and distilled water with 40\% glycerol by volume.
Iodate stock solutions were prepared by dissolving ${\rm KIO_3}$ powder
in distilled water. Arsenous acid stock solutions
were prepared from ${\rm As_2O_3}$ powder. These stock solutions were diluted so that the working 
solution contained [${\rm IO_3^-}$]=0.005M and [As(III)]=0.020M. 
These concentrations were not varied in the experiments we report here. Congo red indicator was used to visualize the reaction fronts. Congo red has a pH range of 3.0 to 5.0, where the acid form is blue and the base form is red. 
The reaction front leaves in its wake a product solution with pH of $\sim 2.7$ so that 
the upwelling blue products are easily visible within the red unreacted solution.

\begin{figure}
\includegraphics[height=5.4cm]{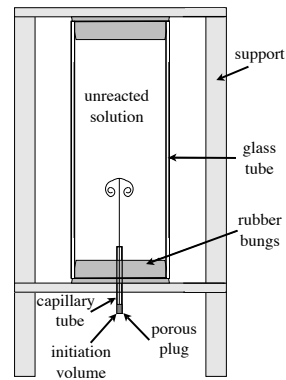}
\caption{\label{schematic}
A schematic of the apparatus. The main volume is a cylinder 32~cm long and  8.9~cm in diameter.  The capillary tube has an inner diameter of 2.7~mm. }
\label{apparatus}
\end{figure}

The apparatus is shown in Fig.~\ref{schematic}.
The reaction tank was a large glass cylinder sealed by large rubber stoppers. A capillary
tube  entered through a hole
in the lower stopper. The outside end of the capillary
tube was sealed with a short rubber tube clamped at one end which formed
the initiation volume. The
rubber tube was filled by a porous plug made of loosely packed cotton.  Reactions
were initiated by inserting a thin needle into the rubber
tube and then injecting a very small amount of product solution
into the plug. The plug served to quench any
hydrodynamic disturbance of the reactant solution during reaction initiation.  The reaction proceeded up the capillary tube, which is large enough that $S > S_c$ and some convection is already present.  It then escapes into the larger tank. 
The apparatus was illuminated from behind and still images of the
evolving front were captured using a digital camera.  The temperature was constant at $24.0 \pm 0.2^\circ$~C.

\begin{figure}
\includegraphics[height=6.2cm]{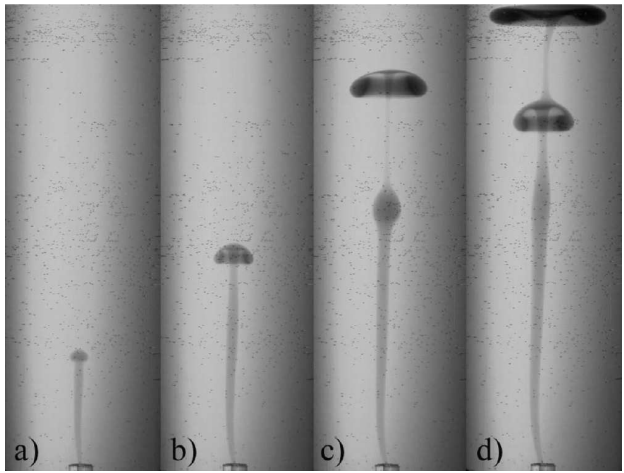}
\caption{\label{plumes}
A sequence of images showing the evolution of the plume structure.  (a) 202~s, (b) 328~s, (c) 458~s and (d) 554~s after release.}
\label{plumes_photo}
\end{figure}

The various stages in the evolution of the resulting free autocatalytic plume are shown in Fig.~\ref{plumes}.
Initially the plume rises out of the capillary tube and its 
head remains roughly spherical and grows slowly. This stage of growth is shown in Fig.~\ref{plumes}a. In this initial stage, there is essentially no entrainment 
of reactant solution into the plume head.

At a well-defined point, entrainment sets in.  Fig.~\ref{plumes}b shows a plume head well past the onset of entrainment. This shows the familiar mushroom-shaped head for a plume in which the surrounding fluid is being drawn into the head by a single overturning vortex ring. 

Once the entraining autocatalytic plume head reaches a second critical size, 
the plume head begins to pinch-off from the upwelling conduit. 
The pinch creates a bottleneck in the conduit which then swells as it fills with 
rising product solution. During the swelling below the bottleneck, what was 
formerly the head of a starting plume becomes an essentially free vortex
ring. As shown in Fig.~\ref{plumes}c, 
the vortex ring detaches from the conduit, and is eventually connected only by a 
very thin filament of product solution. The swelling in the conduit develops into a new, second generation plume head. In Fig.~\ref{plumes}d,  a new pinch-off process has 
started again below the second
generation plume head.  Yet another bottleneck is formed when the second generation
head pinches-off and becomes a vortex ring. The new bottleneck swells and 
the process is repeated. We have observed four generations of pinch-off and 
subsequent vortex ring formation in autocatalytic plumes.   The 
fourth generation plume head eventually reaches the top boundary of
the reaction vessel, closing the possibility of further pinch-off~\cite{movie}.

>From the digital images of plume evolution, we determined the height and width of the 
ascending buoyant plume as a function of time. Height 
was measured from the top of the front to where it exited the capillary 
tube, and the width refers to the maximum width across the head or 
vortex ring.  It is useful to define the dimensionless  ratio $\alpha =  w_h / w_c$, where $w_h$ is the width of the head and $w_c$ is the width of the conduit. Only the first generation head and vortex ring were considered in this analysis.

 We found that the initial stage of growth ends and entrainment begins once 
$\alpha$ reaches a critical value of $\alpha_{c1} = 2.6 \pm 0.1$. Similarly, once 
the entraining plume head reaches a second critical size, $\alpha_{c2}=4.6 \pm 0.1$, 
the plume head begins to pinch-off from the upwelling conduit. 

\begin{figure}
\includegraphics[height=8cm]{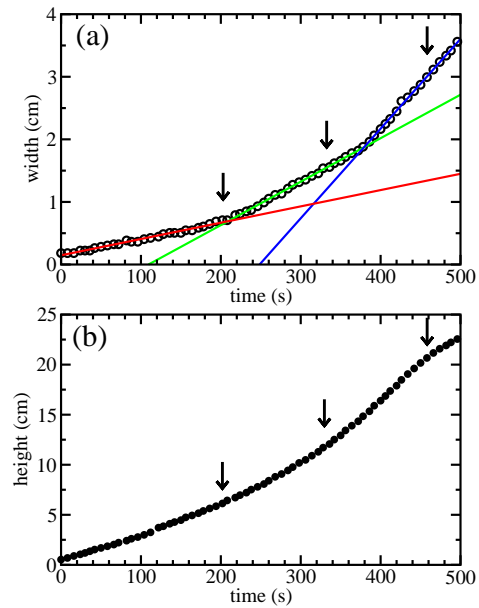}
\caption{\label{plumedata} (Color online) 
The growth of the plume as a function of time, up to the completion of the first pinch-off, showing (a) the width of the head, and (b) its height.  Colored lines in (a) show the nearly linear evolution of the width in the three regimes indicated. Vertical arrows indicate the positions of images (a), (b) and (c) from the previous figure. }
\label{plume_data}
\end{figure}

Typical width and height data for first generation 
head formation of an autocatalytic plume is shown in Fig.~\ref{plumedata}. We observed that the head accelerates during its ascent. The acceleration of the plume head continues 
after it has pinched-off, even though it is almost entirely disconnected 
from the conduit. This acceleration is in contrast to the constant velocity
rise behavior previously observed in conventional plumes~\cite{moses}. The 
continued acceleration of the head once it pinches-off and becomes a thermal 
is also unique to autocatalytic plumes; normal vortex rings expand and slow down after 
they have pinched-off~\cite{turner_01}.

Fig.~\ref{plumedata}a shows that the width of the plume goes through 
three distinct regions of approximately linear growth. 
The initial, and slowest, growth rate obtains when $\alpha < \alpha_{c1}$ and  there is 
no entrainment into the plume head and it remains roughly spherical. 
A transition to a second, increased growth regime occurs
once entrainment begins and $ \alpha_{c1} < \alpha < \alpha_{c2}$. Finally, the third and fastest growth regime commences once $\alpha > \alpha_{c2}$ and the plume head detaches and becomes a thermal. In experiments on conventional plumes 
the growth rate of the head has been determined~\cite{moses} to scale with time $t$ as $t^\beta$, where $\beta=0.54\pm 0.05$. This is clearly not the case for autocatalytic plumes.

%
%
To our knowledge, our experiments mark the first study of buoyant plumes
that are directly driven by an autocatalytic reaction.  Of course, buoyant
plumes~\cite{moses,morton,turner_02,turner_03,griffiths_01} and buoyant
vortex rings~\cite{turner_01,turner_03,lundgren_01} have been the subjects of extensive previous  
study. Refs.~\cite{fay,list,gebhart} review much of the older work.  Some clarification  of nomenclature is useful here.  A {\it plume} is a continuous region of buoyancy supplied by a point source. A {\it starting plume} is a plume with a well-defined, advancing head. These cases of buoyancy driven flow may be distinguished from a {\it thermal}, which is a 
freely evolving buoyant object disconnected from its source. Thermals 
may take the form of vortex rings, similar to smoke rings.

There is great diversity in past experiments on {\it 
passive} plumes, where we use {\it passive} to mean flows that do not react internally to produce buoyancy.  In typical experiments, passive plumes are formed from 
buoyant fluid that has been injected into a less-buoyant medium at a point source. The buoyancy may be due to a temperature or composition difference or both. In 
this case, however, all the buoyancy is derived from processes external to the experiment
and arrives with the volume of injected fluid. Subsequent stirring can only dilute and reduce this buoyancy.  In the case of autocatalytic plumes, the volume of buoyant fluid injected is negligible and all of the buoyancy is generated by the conversion of dense fluid into buoyant fluid at the thin reaction front.  The volume of buoyant fluid is steadily increasing, particularly in the head.

Symmetrical plume heads similar to those we observe can be found  
in passive starting plumes~\cite{griffiths_01}. Moreover, the heads 
of passive buoyant plumes can be observed to pinch-off to form buoyant thermals in the form of vortex  rings. The formation thermals by pinch-off has been the subject of theoretical and 
of experimental interest in passive systems~\cite{lundgren_02,fabris, gharib, shusser}. 
The formation of discrete thermals is a characteristic of systems in
which the ratio of the viscosities of the buoyant fluid
and its surroundings is of order 1~\cite{jellinek}.

There is a critical difference between the passive vortex rings
formed by pinch-off and those that we observe in autocatalytic reactions. As the head 
or vortex ring rises, it entrains surrounding unreacted fluid, stirring it into the
plume head and accelerating the production of new buoyancy.  This feedback mechanism
is unique to autocatalytic plumes.  In our plumes, this stirring is laminar, rather than turbulent, and presumably involves the extreme stretching of the thin reaction front within the vortex ring.

Autocatalytic plumes may have a useful analogy to important geophysical processes.
Conventional plumes have been the subject of numerous laboratory analog experiments
aimed at understanding how material upwelling from deep in the mantle forms oceanic and 
continental hotspots~\cite{griffiths_01}. In addition to plumes, instabilities in the bottom layer 
of the mantle are also likely to give rise to thermals.  Such thermals may incorporate internal sources of buoyancy due to the heat supplied by the breakdown of radionuclides.  Internally heated thermals are one plausible explanation for ocean island volcanism~\cite{griffiths_02}. The buoyancy-producing capabilities of autocatalytic thermals could serve as an interesting laboratory analog of this process.

We have described a study of buoyant, three-dimensional plumes driven by the iodate-arsenous acid reaction. In a water solution that was made more viscous with the
addition of glycerol, we found that a rising plume with a well-defined
head was formed. The plume head grew to a critical size and subsequently
detached from the upwelling conduit to form a buoyant vortex ring. The
entrainment of reactant solution into the autocatalytic chemical plume head
and subsequent vortex ring produced additional buoyancy by assisting the
chemical reaction, leading to an acceleration of the vortex ring. This behavior reverses the usual role of entrainment, which normally acts to reduce buoyancy in conventional passive plumes and thermals.  Autocatalytic chemical smoke rings offer some simple insights into combustion-driven flow and certain geophysical processes.
 
We thank M. Menzinger, A. De Wit, and A. M. Jellinek for helpful discussions and constructive comments. This research was supported by the Natural Science and Engineering Research Council of Canada.

%
%


\begin{references}

\bibitem{combust} J. Warnatz, U. Maas, and R.W. Dibble, {\it Combustion}, 
3rd ed., Springer (2001).

\bibitem{fires} S. R. Tieszen, Ann. Rev. Fluid Mech. {\bf 33}, 67 (2001).
{\bf 84}, 4357 (2000).

\bibitem{hanna} A. Hanna, A. Saul, and K. Showalter, J. Am. Chem. Soc.
{\bf 104}, 3838 (1982).

\bibitem{pojman} J. A. Pojman, I. R. Epstein, T. J. McManus, and
K. Showalter, J. Phys. Chem. {\bf 95}, 1299 (1991).

\bibitem{vasquez_01} D. Vasquez, J. Wilder, and
B. Edwards, Phys. Fluids A {\bf 4}, 2410 (1992).

\bibitem{wilder} J. W. Wilder, D. A. Vasquez and B. F. Edwards,
Phys. Rev. E {\bf 47}, 3761 (1993).

\bibitem{huang} J. Huang, D. A. Vasquez, B. F. Edwards and P. Kolodner,
Phys. Rev. E {\bf 48}, 4378 (1993).

\bibitem{maser} J. Maser, D. A. Vasquez, B. F. Edwards, J. W. Wilder and
K. Showalter, J. Phys. Chem. {\bf 98}, 6505 (1994).

\bibitem{carey} M. R. Carey and S. W. Morris and P. Kolodner,
Phys. Rev. E {\bf 53}, 6012 (1996).

\bibitem{bockmann} M. Bockmann and S. C. M{\"u}ller,
Phys. Rev. Lett. {\bf 85}, 2506 (2000).

\bibitem{dewit} A. De Wit, Phys. Rev. Lett. {\bf 87}, 054502 (2001).

\bibitem{edwards} B. F. Edwards,
Phys. Rev. Lett. {\bf 89}, 104501 (2002).

\bibitem{leconte} M. Leconte, J. Martin, N. Rakotomalala, and D. Salin,
Phys. Rev. Lett. {\bf 90}, 128302 (2003).

\bibitem{mantle} G.F. Davies, M.A. Richards, J. Geology, {\bf100}, 151 (1992).

\bibitem{griffiths_01} R. W. Griffiths, I. H. Campbell, Earth Planet. Sci. 
Lett. {\bf 99}, 66 (1990).

\bibitem{vasquez_02} D. A. Vasquez, B. F. Edwards, and J. W.
Wilder, Phys. Rev A {\bf 43}, 6694 (1991).

\bibitem{front_death} E. Jakab, D. Horv{\'a}th, J. H. Merkin, S. K. Scott, P. L. Simon and A T{\'o}th, Phys. Rev. E., {\bf 66}, 016207 (2002).

\bibitem{movie} For a movie of this process, see\\ http://www.physics.utoronto.ca/nonlinear/movies/plume.html

\bibitem{moses} E. Moses, G. Zocchi, A. Libchaber, J. Fluid. Mech. {\bf 
251}, 581 (1993).

\bibitem{turner_01} J. S. Turner, Proc. R. Soc. Lond. A {\bf 239}, 61 
(1957).

\bibitem{morton} B. R. Morton, G. Taylor, and J. S. Turner, Proc. R. Soc. 
Lond. A
{\bf 234}, 1 (1956).

\bibitem{turner_02} J. S. Turner, J. Fluid. Mech. {\bf 13}, 356 (1962).

\bibitem{turner_03} J. S. Turner, {\it Buoyancy Effects in Fluids},
Cambridge University Press (1973).

\bibitem{lundgren_01} T. S. Lundgren, N. N. Mansour, J. Fluid. Mech. {\bf 224}, 
177 (1991).

\bibitem{fay} J. A. Fay, Ann. Rev. Fluid. Mech., {\bf 5}, 151 (1973).

\bibitem{list} E. J. List, Ann. Rev. Fluid. Mech., {\bf 14}, 189 (1982).

\bibitem{gebhart} B. Gebhart, D. S. Hilder, M. Kelleher, Adv. Heat Transfer
{\bf 16}, 1 (1984).

\bibitem{lundgren_02} T. S. Lundgren, J. Yao, N. N. Mansour, J. Fluid. Mech. {\bf 224}, 
177 (1992).

\bibitem{fabris} D. Fabris and D. Leipmann, Phys. Fluids {\bf 9}, 2801 
(1997).

\bibitem{gharib} M. Gharib, E. Rambod, and
K. Shariff, J. Fluid Mech. {\bf 360}, 121 (1998).

\bibitem{shusser} M. Shusser and M. Gharib, J. Fluid Mech. {\bf 416}, 173 
(2000).

\bibitem{jellinek} A. M. Jellinek, A. Lenardic, M. Manga, Geophys. Res. Lett. {\bf 29}, 10.1029/2001GL014624.

\bibitem{griffiths_02} R. W. Griffiths, Earth Planet. Sci. 
Lett. {\bf 78}, 435 (1986).


\end{references}
\end{document}